\documentclass{article}
\usepackage{graphicx}

\newcommand{\lwig}{\mbox{\,\raisebox{.3ex}
    {$<$}$\!\!\!\!\!$\raisebox{-.9ex}{$\sim$}\,}}
\newcommand{\gwig}{\mbox{\,\raisebox{.3ex}
    {$>$}$\!\!\!\!\!$\raisebox{-.9ex}{$\sim$}}\,}

\newif\ifhepph

\hepphtrue

\usepackage{fortschritte}
\def\beq{\begin{equation}}                     %
\def\eeq{\end{equation}}                       %
\def\bea{\begin{eqnarray}}                     
\def\eea{\end{eqnarray}}                       
\begin {document}                 
\def\titleline{
\ifhepph
{\rm\normalsize\rightline{DESY 02-159}\rightline{hep-ph/0212342}}
\vskip 1cm 
\fi
%
%
%
Production of Black Holes in TeV-Scale Gravity\ifhepph\footnote[1]{Talk presented at 
the 35$^{\rm th}$ International Symposium Ahrenshoop on the Theory of Elementary Particles, 
Aug. 26-30, 2002, Berlin-Schm\"ockwitz, Germany.}\fi 
}
\def\email_speaker{
{\tt 
%
%
andreas.ringwald@desy.de
}}
\def\authors{
%
%
%
%
%
A. Ringwald
}
\def\addresses{
%
%
%
%
Deutsches Elektronen-Synchrotron DESY, Notkestra\ss e 85, D-22607 Hamburg, Germany\\
}
\def\abstracttext{
%
%
Copious production of microscopic black holes is one of the least model-dependent
predictions of TeV-scale gravity scenarios. 
We review the arguments behind this assertion and 
discuss opportunities to track the 
striking associated signatures in the near future. These include searches
at neutrino telescopes, such as AMANDA and RICE,  
at cosmic ray air shower facilities, such as the Pierre Auger Observatory,  
and at colliders, such as 
the Large Hadron Collider.                         
}
\large
\makefront
\section{Introduction}

Black holes are among the most remarkable, but also most mysterious objects in physics. 
Since Hawking's prediction of black hole evaporation~\cite{Hawking:rv}, 
they play an important role in any attempt towards a theory of quantum gravity.
Unfortunately, experimental detection of Hawking radiation from real, massive ($M_{\rm bh}$) 
astrophysical black holes seems impossible, 
since the corresponding temperature, as seen by an outside stationary observer, is tiny,
$T_{\rm H} = 6\cdot 10^{-8}\ {\rm K}\ \left( M_\odot/M_{\rm bh}\right)$. 
Furthermore, the proposed detection~\cite{Halzen:uw} of Hawking radiation 
from primordial mini black holes -- possible relics from the big bang -- would be rather 
indirect. Last, but not least, the production and exploration of microscopic black holes 
at terrestrial accelerators requires seemingly center-of-mass (cm) energies of order the Planck
scale, $\sqrt{\hat s}\,\gwig\, \bar M_{\rm Pl}=1/\sqrt{8\pi G_N}=2\cdot 10^{18}$~GeV, way beyond energies obtainable 
even in distant future~\cite{'tHooft:1987rb}.   

The latter, so far remote possibility seems, however, meanwhile within reach in the context of 
theories with $\delta = D-4\geq 1$ 
flat~\cite{Arkani-Hamed:1998rs} or warped~\cite{Randall:1999ee} extra dimensions 
and a low fundamental Planck scale 
$M_D\,\gwig$ TeV characterizing quantum 
gravity.
In these theories one expects the copious production of black holes 
in high energy collisions
at cm energies above 
$M_D$~\cite{Argyres:1998qn}.
Correspondingly, the Large Hadron Collider (LHC)\ifhepph~\cite{Evans:2001mn}\fi, 
expected  to have a first physics run in  2006, may turn into a factory of black holes
at which their production and evaporation may be studied in 
detail~\cite{Giddings:2001bu,Dimopoulos:2001hw,Dimopoulos:2001qe}. 
But even before the commissioning of the LHC, the first signs of black
hole production may be observed in the scattering of  ultrahigh energy 
cosmic neutrino off nuclei in ice or air~\cite{Feng:2002ib,Emparan:2001kf,Ringwald:2002vk,Anchordoqui:2001cg,%
Kowalski:2002gb,Alvarez-Muniz:2002ga} and recorded by 
existing neutrino telescopes, such as AMANDA\ifhepph~\cite{Andres:2001ty}\fi\ and 
RICE\ifhepph~\cite{Kravchenko:2001id}\fi, or
at cosmic ray air shower facilities, such as the Pierre Auger 
Observatory\ifhepph~\cite{Zavrtanik:2000zi}\fi. 
Moreover, already now sensible constraints
on black hole production can be obtained~\cite{Ringwald:2002vk,Anchordoqui:2001cg} from the non-observation of horizontal 
showers by the Fly's Eye collaboration~\cite{Baltrusaitis:1985mt} and the 
Akeno Giant Air Shower Array (AGASA) collaboration~\cite{Yoshida:2001}, 
respectively.  These constraints turn out to be 
competitive with other currently available constraints on TeV-scale gravity 
which are mainly based on interactions associated with
Kaluza-Klein gravitons, 
according to which a fundamental Planck scale as low as $M_D = {\mathcal O}(1)$ TeV is still allowed for $\delta\geq 6$ flat 
or $\delta\geq 1$ warped extra dimensions~\cite{Peskin:2000ti}\ifhepph\footnote{For an exhaustive list of references in this 
context, see Ref.~\cite{Ringwald:2002vk}.}\fi.

In this talk we shortly review the theory and phenomenology of black hole production
in TeV-scale gravity scenarios. More details and an exhaustive reference list can be
found, e.g., in Ref.~\cite{Tu:2002xs}. 

\begin{figure}
\begin{center}
\vspace{-0.5cm}
\parbox{7cm}{\vspace{-6cm}\includegraphics*[width=7cm]{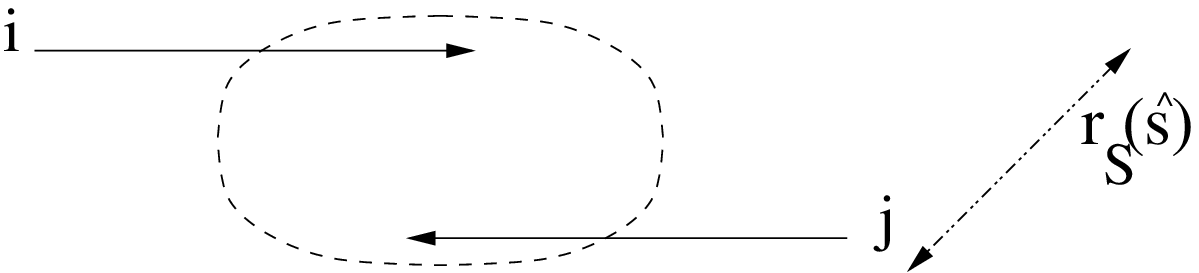}}
\hfill
\includegraphics*[width=5.7cm]{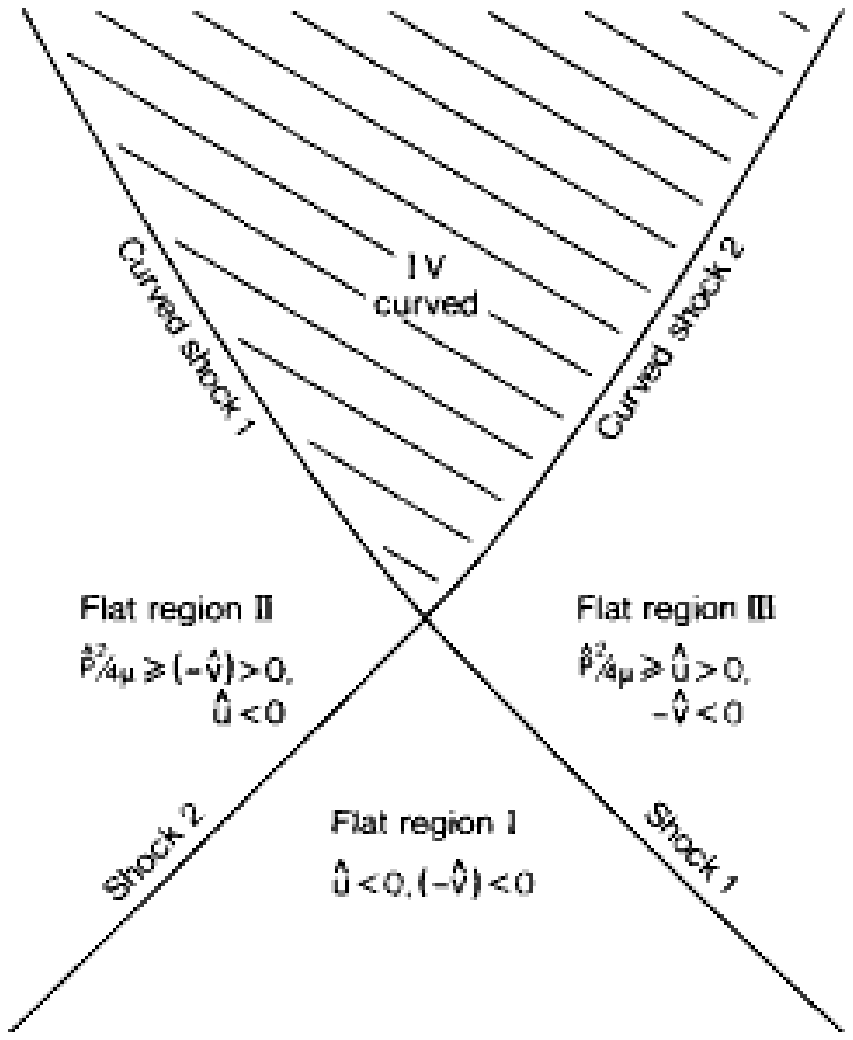}
\caption[dum]{\label{trans_reg}
{\em Left:} Illustration of trans-Planckian scattering of two partons (i,j) at small
impact parameter $b< r_{\rm S}$ (adapted from Ref.~\cite{Giddings:2001bu}).
{\em Right:} Schematic space-time diagram for a trans-Planckian collision~\cite{D'Eath:hb}.  
}
\end{center}
\end{figure}

\section{Black Hole Production -- Theory}

Why one expects the copious production of black holes
in trans-Planckian scattering at small impact parameters? This expectation is 
basically relying on two assertions, which are substantiated below: {\em i)} Trans-Planckian scattering is 
{\em semi-classical}. {\em ii)} There are {\em classical scattering solutions}
corresponding to the formation of black holes at trans-Planckian energies 
$\sqrt{\hat s}\gg M_D$ and at small impact parameters $b\,\lwig\,r_{\rm S}(\sqrt{\hat s})$, 
where $r_{\rm S}(M_{\rm bh})$ is the Schwarzschild radius~\cite{Myers:1986un} of a black hole with mass $M_{\rm bh}$
(cf. Fig.~\ref{trans_reg} (left)). Correspondingly, one expects a geometric cross-section
for black hole formation~\cite{Giddings:2001bu,Dimopoulos:2001hw},
\begin{eqnarray}
\label{cross-geom}
     \hat\sigma_{ij}^{\rm bh}\approx  \pi\,r_{\rm S}^2
     \left( \sqrt{\hat s}\right) 
= 
\frac{\pi}{M_D^2}
       \left[
       \frac{\sqrt{\hat s}}{M_D}
       \left(
       \frac{2^\delta \pi^{\frac{\delta -3}{2}}\,\Gamma\left( \frac{3+\delta}{2}\right)}{2+\delta}
       \right)
       \right]^{\frac{2}{1+\delta}}
\,,
\hspace{4ex}
{\rm at}\ \sqrt{\hat s}\gg M_D\,,
\end{eqnarray}     
whose scale is set essentially by the fundamental Planck scale $M_D\,\gwig$~TeV and therefore 
quite sizeable.  

{\em Ad i)} 
The fact that trans-Planckian scattering is semi-classical can be most easily seen through a 
dimensional analysis, in which one keeps $c=1$, but restores the appropriate powers of $\hbar\not= 1$. 
Relevant quantities to be considered are the fundamental Planck scale $M_D$ and the 
fundamental Planck length $\lambda_D$ -- the length below which quantum gravity fluctuations 
of the geometry are important -- in terms of the $D=4+\delta$ dimensional gravitational constant $G_D$,
\begin{eqnarray}
\label{MD-lD-GD}
{M}^{2+\delta}_{D} = \frac{(2\pi)^{\delta -1}}{4}\frac{\hbar^{1+\delta}}{ G_D}, \hspace{6ex}
{ \lambda}^{2+\delta}_{ D} = \hbar\,{ G_D}\,, 
\end{eqnarray}
as well as the de Broglie wave length $\lambda_{\rm B}$ of the scattering quanta and the 
Schwarzschild radius $r_{\rm S}$,
\begin{eqnarray}
\label{lB-rS}
{\lambda_{\rm B}} = 4\pi \frac{\hbar}{\sqrt{\hat s}}\,, \hspace{6ex}
       {r_{\rm S}} =\frac{1}{\sqrt{\pi}}
       \left[
       \frac{
       8\,\Gamma\left( \frac{3+\delta}{2}\right)}{2+\delta}
       \right]^{\frac{1}{1+\delta}}
       \left( {G_D}\,\sqrt{\hat s}\right)^{\frac{1}{1+\delta}}
\,.
\end{eqnarray}
An inspection of Eqs.~(\ref{MD-lD-GD}) and (\ref{lB-rS}) reveals that in the semi-classical 
($\hbar\to 0$) limit, with $G_D$ and $\sqrt{\hat s}$ fixed, the quantities 
$M_D,\lambda_D,\lambda_{\rm B}\to 0$, whereas $r_{\rm S}$ remains 
finite. Therefore, semi-classics corresponds to the trans-Planckian regime, 
$\sqrt{\hat s}\gg M_D$. Moreover, in this regime, the 
Schwarschild radius $r_{\rm S}$ $(\gg\lambda_D\gg \lambda_{\rm B})$ characterizes the 
dynamics.  

\begin{figure}
\begin{center}
\vspace{-0.2cm}
\includegraphics*[bbllx=20pt,bblly=221pt,bburx=570pt,bbury=608pt,width=7.9cm]{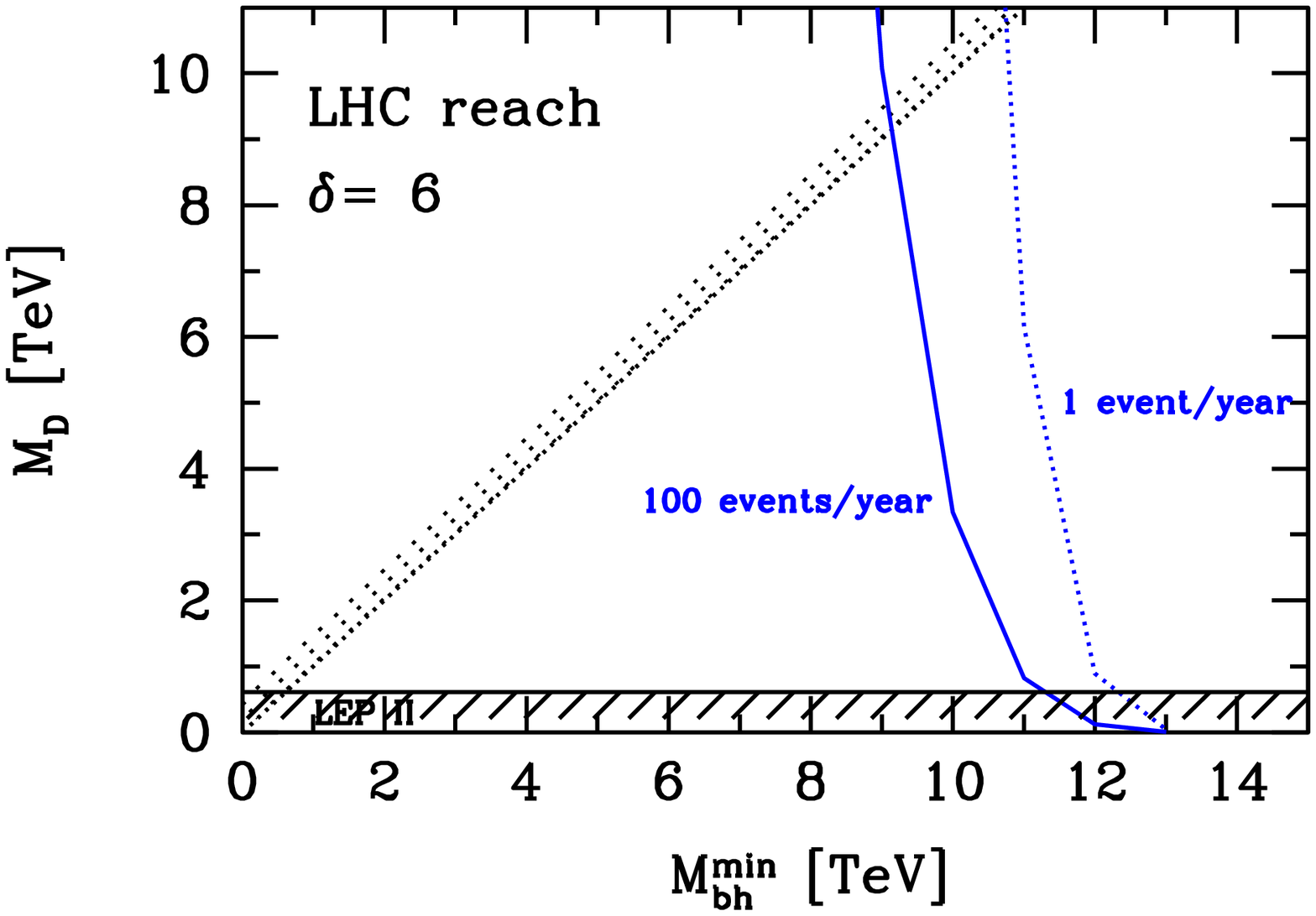}
\hspace{1.55cm}
\parbox{5.cm}{\vspace{-6cm}\includegraphics*[width=4.9cm]{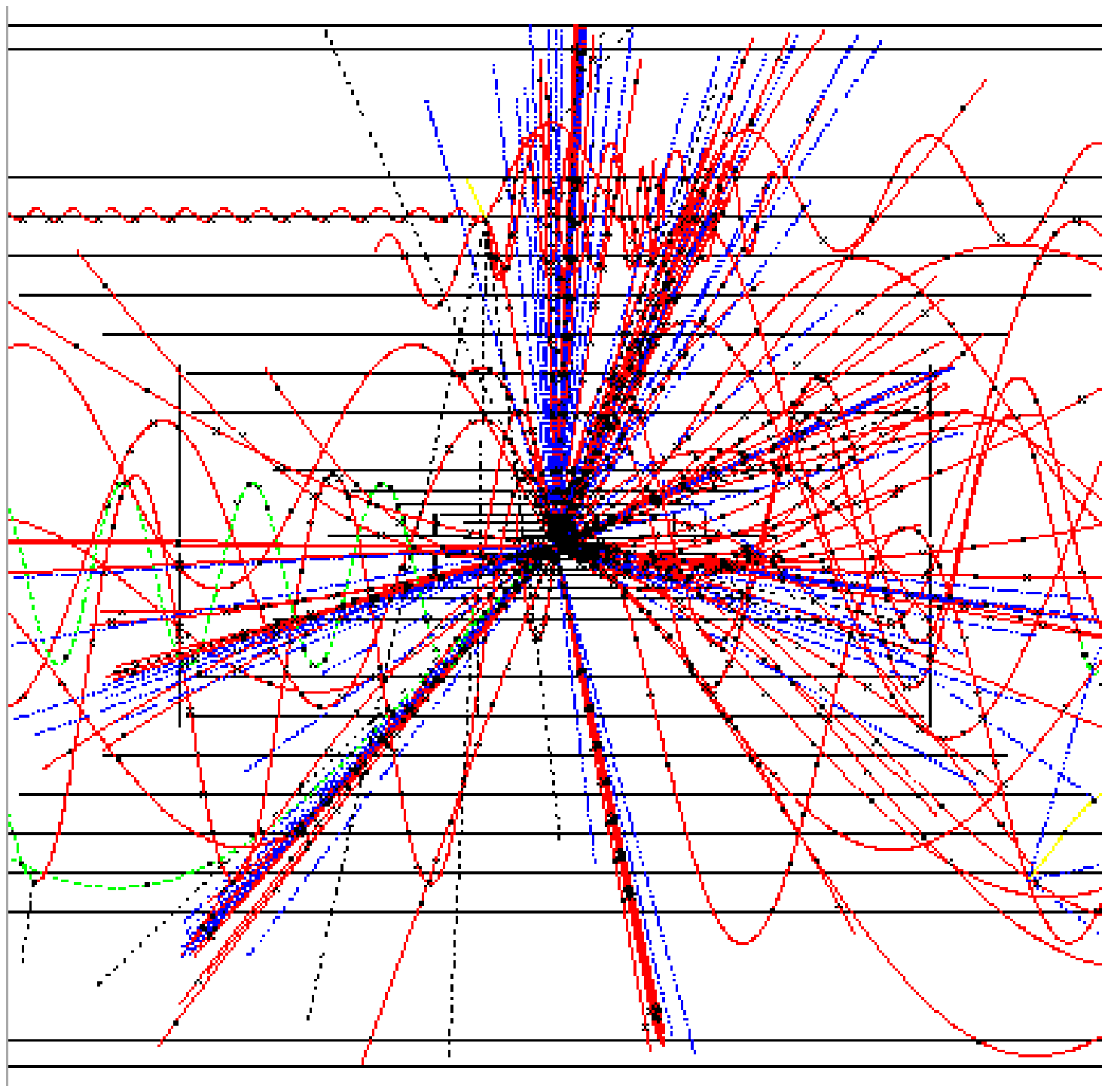}}
\caption[dum]{\label{illu-shock-lhc}
{\em Left:} Accessible region in the black hole production parameters at the LHC for $\delta =6$ extra dimensions 
(adapted from Ref.~\cite{Ringwald:2002vk}).  
The solid and the dotted lines are 
contours of constant numbers of produced black holes per year ($10^7$ s) with a mass larger than $M_{\rm bh}^{\rm min}$, 
for a fundamental Planck mass $M_D$. The shaded dotted, $M_D = M_{\rm bh}^{\rm min}$, 
line gives a rough indication of the boundary of applicability of  the semi-classical 
picture. The shaded solid line gives the constraint $M_D > 0.61$~TeV ($\delta = 6$) from LEP II searches~\cite{Peskin:2000ti}. 
{\em Right:} Black hole final state in a detector simulation~\cite{Barklow:2001mm}.
}
\end{center}
\end{figure}

{\em Ad ii)}
There are basically two regimes in trans-Planckian scattering: The soft regime of large impact 
parameters, $b\gg r_{\rm S}(\sqrt{\hat s})$, which is adequately described by semi-classical eikonal methods for 
elastic small angle ($\theta\sim (r_{\rm S}/b)^{\delta +1}\ll 1$) scattering~\cite{'tHooft:1987rb,Giudice:2001ce}, 
and the hard regime of small impact parameters, $b\ll r_{\rm S}(\sqrt{\hat s})$, in which one expects the gravitational
collapse to a black hole, since at particle crossing an amount $\sqrt{\hat s}$ of energy is localized
within a radius $b<r_{\rm S}$ (cf. Fig.~\ref{trans_reg} (left)). This is basically a variant of Thorne's hoop 
conjecture~\cite{Thorne:hoop} in 
classical general relativity, 
according to which a horizon forms if and only if  a mass $M$ is compacted into a region
      whose circumference in every direction is less than $2\pi r_{\rm S}(M)$.
Indeed, for small enough impact parameter $b<b_{\rm max}$, a marginally trapped 
surface forms at the overlap between two gravitational shockwaves describing the incoming
partons at trans-Planckian energies (cf. Fig.~\ref{trans_reg} (right)), as has been convincingly demonstrated in 
Refs.~\cite{Penrose:unp,D'Eath:hb,Eardley:2002re}. Correspondingly, a geometrical 
cross-section $\approx\pi\, b_{\rm max}^2$ is expected, and a lower bound on $b_{\rm max}$ 
leads to a lower limit of e.g.  
\begin{eqnarray}
     \hat\sigma^{\rm bh}_{ij}\gwig  0.65\,\pi\,r_{\rm S}^2 (\sqrt{\hat s})\,,\hspace{6ex}
     M_{\rm bh}\gwig 0.5\,\sqrt{\hat s}\,,
\end{eqnarray}  
in the case of $D=4$, substantiating the estimate~(\ref{cross-geom})\ifhepph\ (see Ref.~\cite{Voloshin:2001vs} 
for further discussion)\fi. 

\begin{figure}
\vspace{-0.2cm}
\begin{center}
\includegraphics*[bbllx=20pt,bblly=221pt,bburx=570pt,bbury=608pt,width=7.9cm]{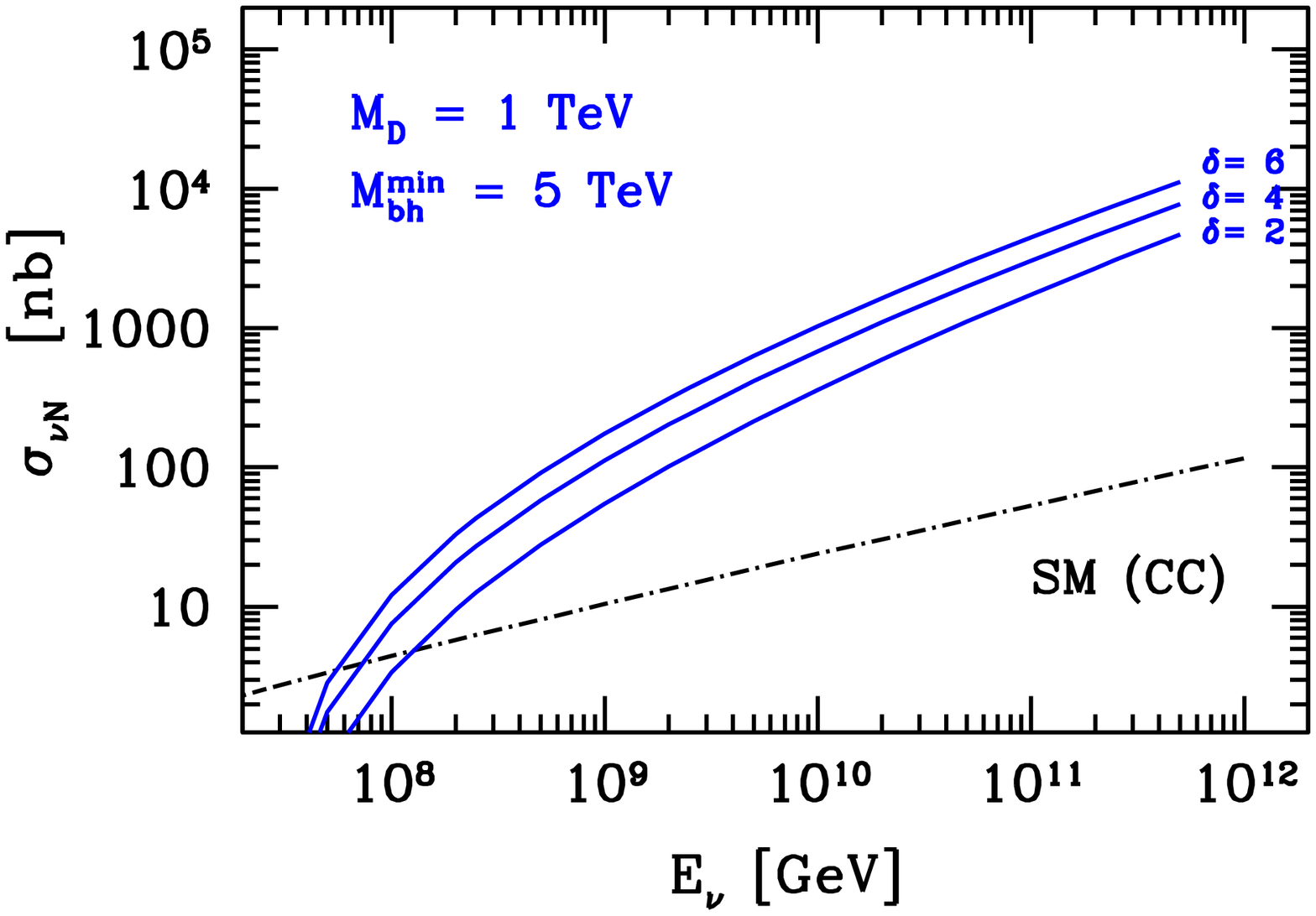}
\hfill
\includegraphics*[bbllx=20pt,bblly=221pt,bburx=570pt,bbury=608pt,width=7.9cm]{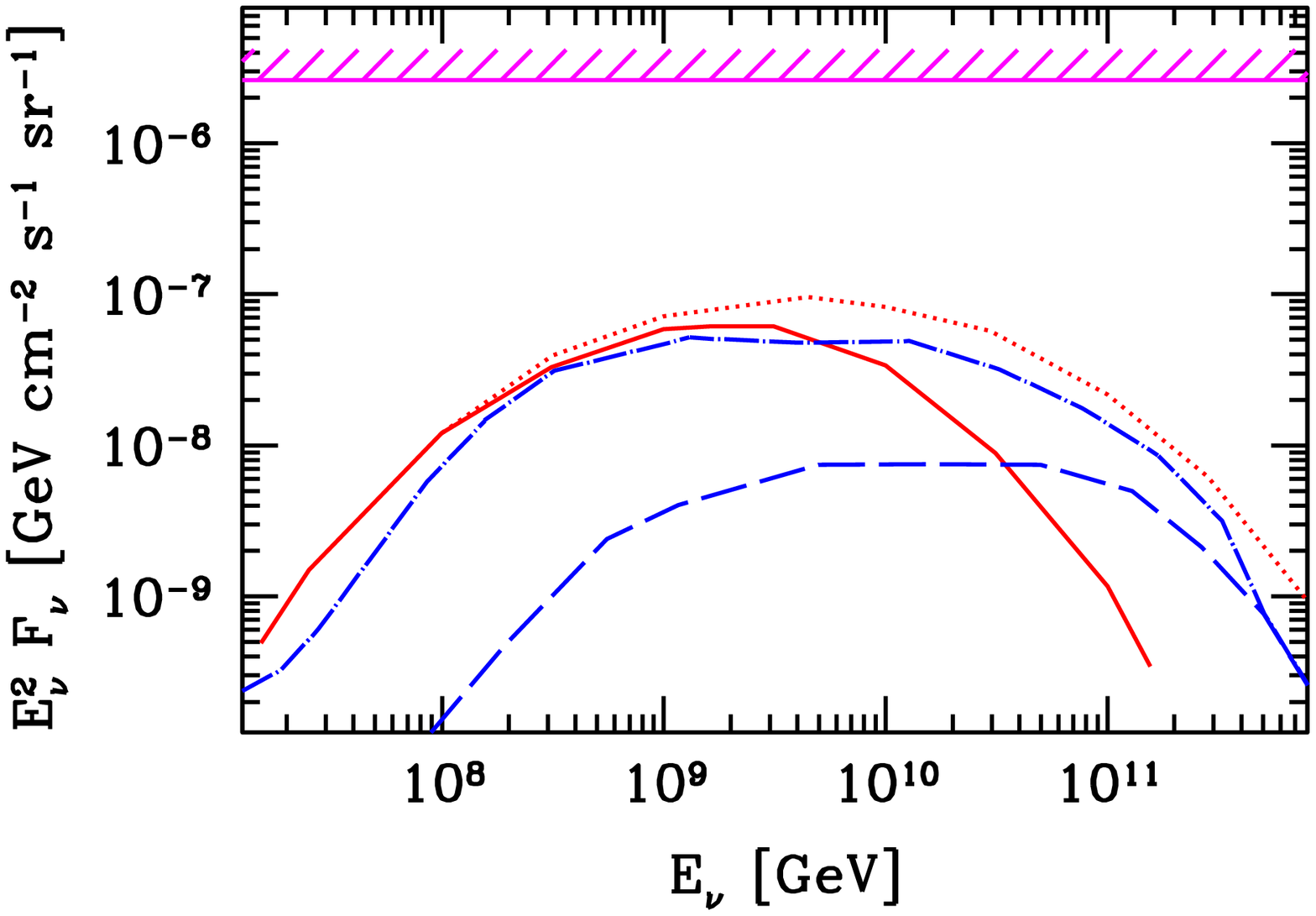}
\caption[dum]{\label{nuN_cross}
{\em Left:} Cross section for black hole production (solid) and 
Standard Model (SM) charged current processes (dashed-dotted) 
in neutrino-nucleon scattering~\cite{Ringwald:2002vk}.
{\em Right:} Predictions of the cosmogenic neutrino flux from Ref.~\cite{Yoshida:1993pt} 
(long-dashed and long-dashed--dotted) and Ref.~\cite{Protheroe:1996ft} 
(solid and dotted).
Shaded: Theoretical upper limit of the neutrino flux from ``hidden'' 
astrophysical sources that are
non-transparent to ultrahigh energy nucleons~\cite{Mannheim:2001wp}.
}
\end{center}
\end{figure}

\section{Black Hole Production -- Phenomenology}

The reach of the LHC to black hole production is illustrated in Fig.~\ref{illu-shock-lhc} (left) 
for $\delta = 6$ extra dimensions.   
As can be seen, the LHC can explore the production of black holes
with minimum masses nearly up to its kinematical limit of $14$ TeV.
In order to appreciate the event numbers indicated in Fig.~\ref{illu-shock-lhc} (left), let us mention the 
expected signature of black hole decay, which is quite spectacular. Once produced, black holes decay primarily 
via Hawking radiation~\cite{Hawking:rv} into a large number of ${\mathcal O}(20)$ hard quanta, with energies 
approaching several hundreds of GeV. A substantial fraction of the beam energy is deposited in visible
transverse energy, in an event with high sphericity (cf. Fig.~\ref{illu-shock-lhc} (right)). 
From previous studies of electroweak sphaleron production~\cite{Aoyama:1986ej}, which has quite 
similar event characteristics~\cite{Farrar:1990vb}, as well as from first 
event simulations 
of black hole production~\cite{Dimopoulos:2001hw}, it is clear that only a handful of 
such events is needed at the LHC to discriminate them from perturbative Standard Model background.

\begin{figure}
\begin{center}
\includegraphics*[bbllx=20pt,bblly=221pt,bburx=570pt,bbury=608pt,width=7.9cm]{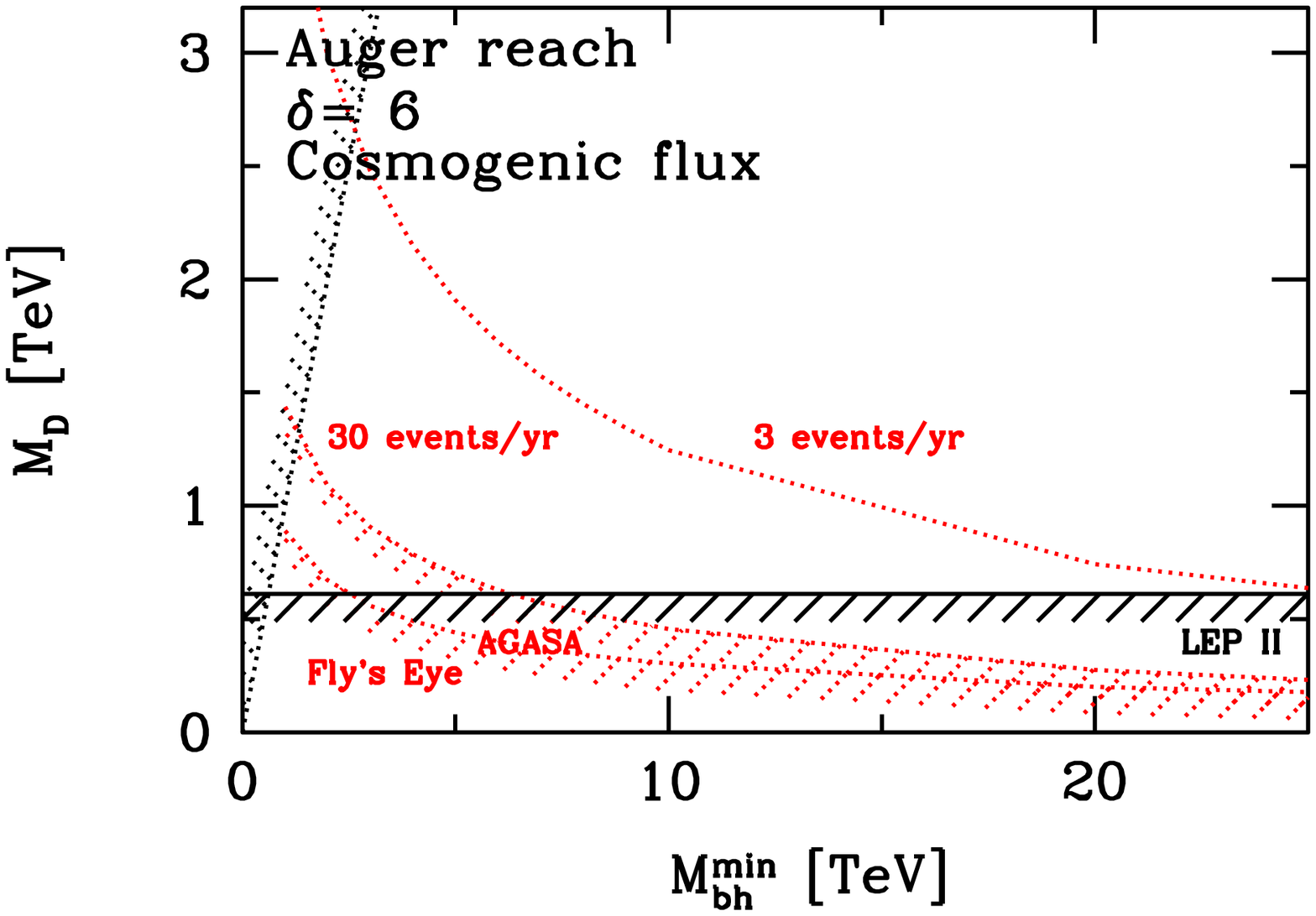}
\hfill
\includegraphics*[bbllx=20pt,bblly=221pt,bburx=570pt,bbury=608pt,width=7.9cm]{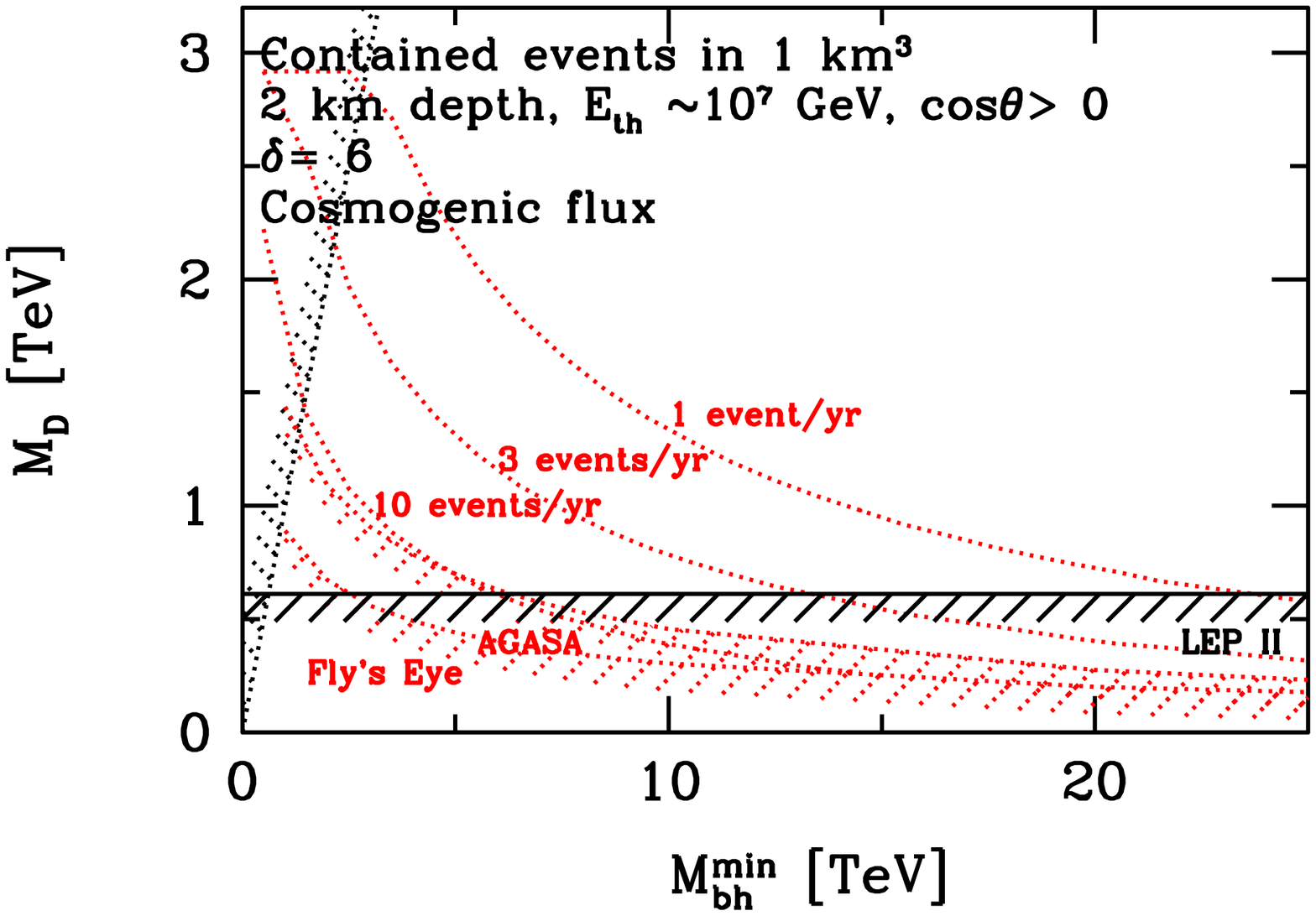}
\caption[dum]{\label{auger-nt-casc}
{\em Left:} Projected Auger reach in the black hole production parameters for $\delta = 6$ extra 
dimensions~\cite{Ringwald:2002vk},  
exploiting the 
cosmogenic neutrino flux from Fig.~\ref{nuN_cross} (right) ($1\ {\rm yr}=3.16\cdot 10^7$~s).
The shaded dotted line labeled ``Fly's Eye'' (``AGASA'') indicates the constraint arising from the non-observation of 
horizontal showers by the Fly's Eye collaboration~\cite{Baltrusaitis:1985mt,Ringwald:2002vk} 
(AGASA~\cite{Yoshida:2001,Anchordoqui:2001cg}). 
{\em Right:} Reach in the black hole production parameters for $\delta = 6$ extra dimensions, 
for contained events in an under-ice detector at a depth of 2 km and with an 1 km$^3$ 
ficudial volume, exploiting the 
cosmogenic neutrino flux from Fig.~\ref{nuN_cross} (right)~\cite{Kowalski:2002gb}.

}
\end{center}
\end{figure}

However, until the LHC starts operating, cosmic rays provide the only 
access to the required energy scales.
Cosmic rays of energies up to $\simeq 10^{21}$ eV have been observed.
The ``cosmogenic'' neutrinos, expected from the cosmic ray interactions 
with the cosmic microwave background (e.g. $p \gamma \rightarrow \Delta 
\rightarrow n \pi^+ \rightarrow \nu_{\mu} \bar{\nu}_{\mu} \nu_e...$), 
are more or less guaranteed to exist among ultrahigh energy cosmic 
neutrinos predicted from various sources\ifhepph\ (cf. Ref.~\cite{Kalashev:2002kx} for 
a most recent discussion)\fi.
Thus, if TeV-scale gravity is realised in nature, ultrahigh energy cosmic rays/cosmic 
neutrinos should have been producing mini black holes in the atmosphere 
throughout earth's history. 
For cosmic ray facilities such as Fly's Eye, AGASA and Auger, the clearest 
black hole signals are neutrino-induced quasi-horizontal air showers which 
occur at rates exceeding the Standard Model rate by a factor of $10 \div 10^2$ 
(see Fig.~\ref{nuN_cross} (left)), and have distinct characteristics 
\cite{Feng:2002ib,Emparan:2001kf,Ringwald:2002vk,Anchordoqui:2001cg}. 
Black hole production could also enhance the detection rate at neutrino 
telescopes such as AMANDA/IceCube, ANTARES, Baikal, and RICE significantly, 
both of contained and of through-going events~\cite{Kowalski:2002gb,Alvarez-Muniz:2002ga}. 

The reach of cosmic ray facilites in black hole production has been 
thoroughly investigated  by
exploiting the cosmogenic neutrino fluxes of Fig.~\ref{nuN_cross} 
(right)~\cite{Ringwald:2002vk,Anchordoqui:2001cg}.
It is argued in Ref.~\cite{Anchordoqui:2001cg} that an excess of a handful
of quasi-horizontal events are sufficient for a discrimination against the
Standard Model background.
An inspection of Fig.~\ref{auger-nt-casc} (left) thus leads to the conclusion
that the Pierre Auger Observatory, expected to become fully operational in 
2003, has the opportunity to see first signs of black hole production. 
Moreover, as also shown in Fig.~\ref{auger-nt-casc} (left), the non-observation 
of horizontal air showers reported by the Fly's Eye~\cite{Baltrusaitis:1985mt} and the 
AGASA~\cite{Yoshida:2001} collaboration provides a stringent bound on 
$M_D$, which is competitive with existing bounds on $M_D$ from colliders 
as well as from astrophysical and cosmological considerations, particularly 
for larger numbers of extra dimensions ($\delta \geq 5$) and smaller 
threshold ($M^{\rm min}_{\rm bh} \lwig 10$ TeV) for the semi-classical 
description. 

As for neutrino telescopes, investigations show~\cite{Kowalski:2002gb,Alvarez-Muniz:2002ga} 
(cf. Fig.~\ref{auger-nt-casc}
(right))  that due to their small volume $V \approx 0.001 \div 0.01$ km$^3$ 
for contained events, the currently operating 
neutrino 
telescopes AMANDA and Baikal cannot yield a large enough contained event
rate to challenge the already existing limits from Fly's Eye and AGASA.
Even IceCube does not seem to be really competitive, since the final 
effective volume $V \approx 1$ km$^3$ will be reached only after the LHC
starts operating and Auger has taken data for already a few years.
But sensible constraints on black hole production can be expected from RICE, 
a currently operating radio-Cherenkov neutrino detector with an effective 
volume $\approx 1$ km$^3$ for $10^8$ GeV electromagnetic cascades, using 
already availabe data. 

\begin{figure}
\vspace{-0.25cm}
\begin{center}
\includegraphics*[bbllx=20pt,bblly=221pt,bburx=570pt,bbury=608pt,width=7.9cm]{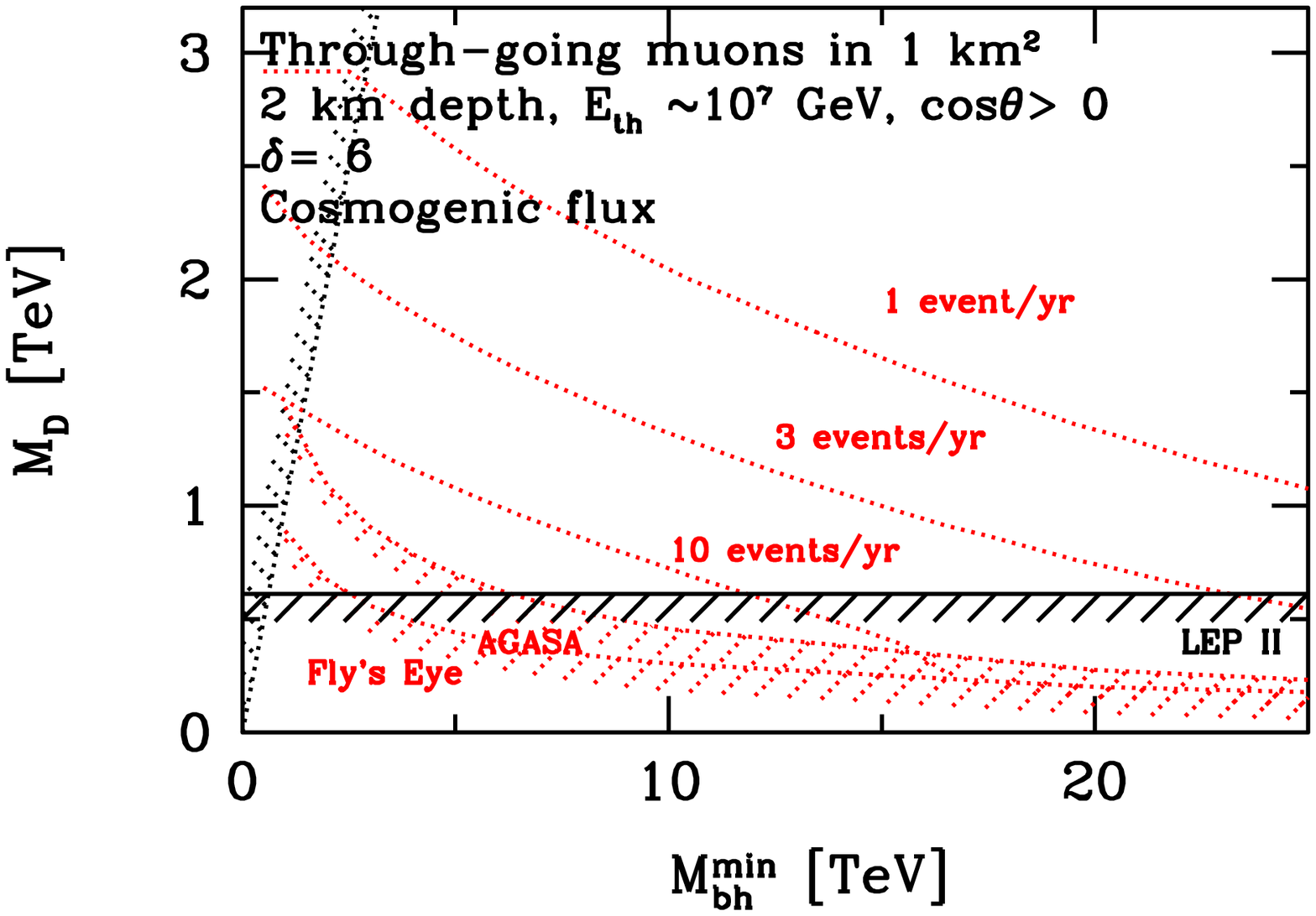}
\hfill
\includegraphics*[bbllx=20pt,bblly=221pt,bburx=570pt,bbury=608pt,width=7.9cm]{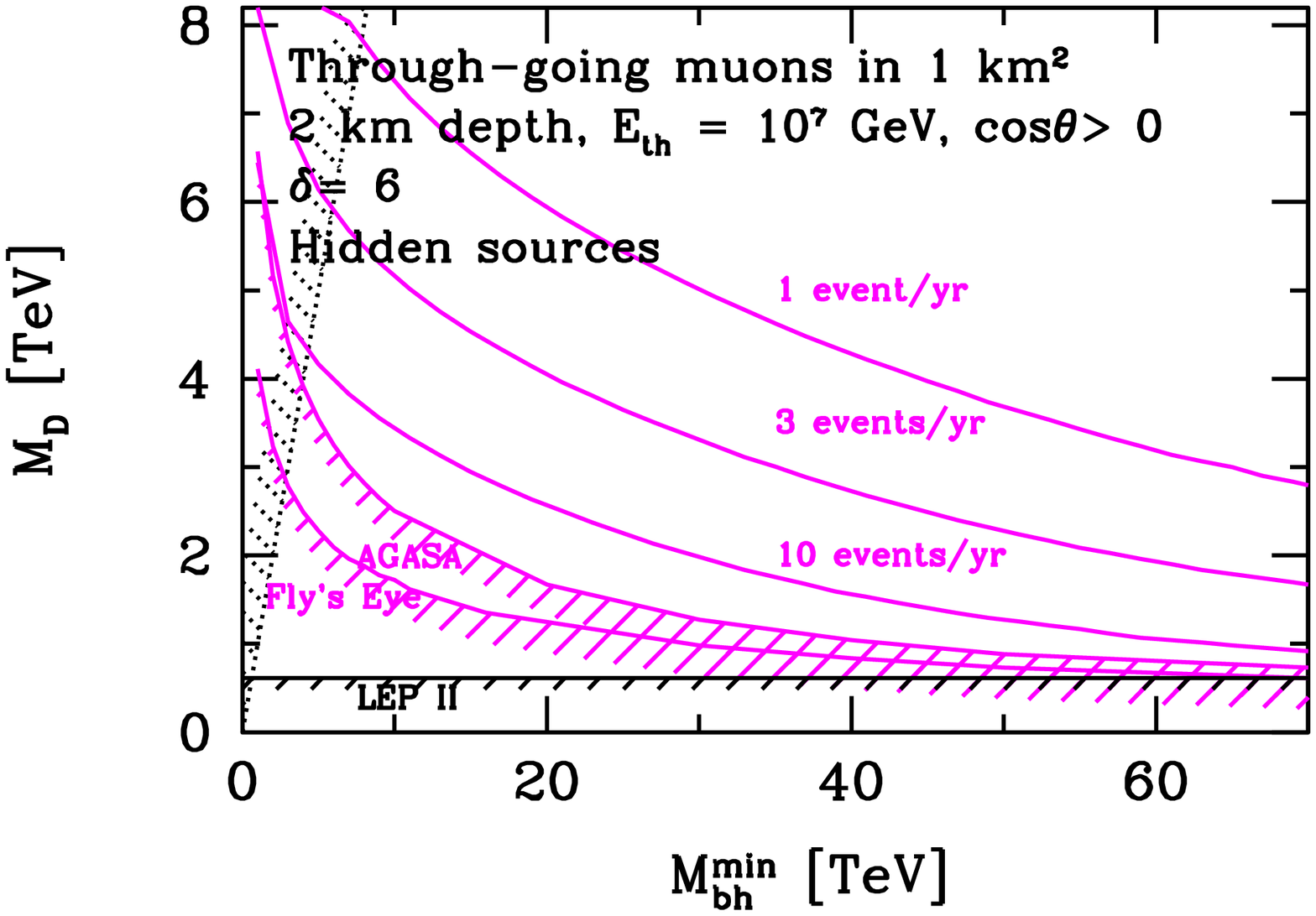}
\caption[dum]{\label{nt-muons}
Reach in the black hole production parameters for $\delta = 6$ extra dimensions, 
for through-going muons in an under-ice detector at a depth of 2 km and with an 1 km$^2$ 
effective area ($1\ {\rm yr}=3.16\cdot 10^7$~s)~\cite{Kowalski:2002gb}. 
{\em Left:} Exploiting the 
cosmogenic neutrino flux from Fig.~\ref{nuN_cross} (right).
{\em Right:} Exploiting 
the upper limit on the neutrino flux from ``hidden'' hadronic astrophysical sources from
Fig.~\ref{nuN_cross} (right). 
}
\end{center}
\end{figure}

The ability to detect muons from distant neutrino reactions increases an
underwater/-ice detector's effective neutrino target volume dramatically.
In the case that the neutrino flux is at the level of the cosmogenic one,
only  $\lwig 1$ events from Standard Model background are
expected per year.
Thus, with an effective area of about 0.3 km$^2$ for down-going muons above
$10^7$ GeV and 5 years data available, AMANDA should be able to impose 
strong constraints if no through-going muons above $10^7$ GeV are seen in
the currently available data (cf. Fig.~\ref{nt-muons} (left)).
Moreover, in the optimistic case that an ultrahigh energy cosmic neutrino 
flux significantly higher than the cosmogenic one is realised in nature,
one even has discovery potential for black holes at IceCube beyond the 
reach of LHC, though discrimination between Standard Model background
and black hole events becomes crucial (cf. Fig.~\ref{nt-muons} (right)). 

\mbox{}
\\ \\  
{\bf Acknowledgement} I would like to thank M. Kowalski and H. Tu 
for the nice collaboration. 

\end{document}